# Broken Letters, Broken Narratives:
# A Case Study on Arabic Script in DALL-E 3

**Arshia Sobhan[1], Philippe Pasquier[2], Gabriela Aceves Sepúlveda[3]**

[1,2,3]Simon Fraser University
Vancouver, British Columbia, Canada
[1]asobhans@sfu.ca, [2]pasquier@sfu.ca, [3]gabriela_aceves-sepulveda@sfu.ca

**Abstract**

Text-to-image generative AI systems exhibit significant limitations when engaging with under-represented domains, including non-Western art forms, often perpetuating biases and misrepresentations. We present a focused case study on the generative AI system DALL-E 3, examining its inability to properly represent calligraphic Arabic script, a culturally significant art form. Through a critical analysis of the generated outputs, we explore these limitations, emerging biases, and the broader implications in light of Edward Said's concept of Orientalism as well as historical examples of pseudo-Arabic. We discuss how misrepresentations persist in new technological contexts and what consequences they may have.

## Introduction

In recent years, text-to-image generative AI has rapidly expanded into various artistic domains, bringing both creative potential and challenges. While these advancements offer new possibilities for artistic expression, they also reveal significant limitations when applied to under-represented domains, including non-Western art forms—under-represented historically in media, in the training datasets used to develop these systems, and as part of the problem definitions guiding generative AI development and modification. The development of mainstream generative AI systems, particularly those developed based on big data, inherits biases ingrained in datasets and is influenced by design decisions often rooted in Western visual culture. Consequently, these systems may struggle to accurately represent culturally sensitive art forms outside of this context.

The motivations behind this study stem from a need to better understand how generative AI technologies engage with non-Western art forms and to examine the effects on culturally significant art forms. This paper investigates the implications of this dynamic through a case study focused on the representation of Arabic script in text-to-image generative AI. As a writing system, this script is shared among multiple languages, and its calligraphic form is a prominent art form deeply embedded in the cultural fabric of societies that use the script.

The case study examines the treatment of Arabic script by the widely used text-to-image generative AI system DALL-E 3, analyzing a corpus of generated images to assess how the script is represented. Through a critical analysis of the corpus and generative processes, this study highlights the implications of these representations. By drawing parallels to other known phenomena related to Arabic script, the study aims to clarify the extent of issues presented by generative AI technology and to better understand the consequences in light of past instances of mistreatment and misrepresentation of the script.

This study is not intended as a technical evaluation of the specific generative AI system employed. Instead, it seeks to clearly illustrate a failure case and critically engage with its implications. Hence, The focus is on the broader significance of this issue for a non-Western art form, rather than a detailed technical investigation of the system itself.

Recognizing the vast diversity of non-Western art practices, each shaped by unique cultural, aesthetic, and historical contexts, this paper does not attempt to generalize its findings across all non-Western domains. Rather, it focuses on the real-world impact of generative AI on a specific domain. It is hoped that this study will provide insights to encourage and guide future investigations into the challenges and implications of generative AI within other non-Western contexts.

It is also acknowledged that readers unfamiliar with the Arabic script and its calligraphic nuances may find it challenging to visually assess the results presented in this paper. This limitation is inevitable given the rich cultural knowledge embedded within the script's aesthetic dimensions. The observations and analysis in this study draw on the domain expertise of the primary author who is native to the script with professional experience in calligraphic Arabic.

## Background

Arabic script, the world's second most-used alphabetic writing system, is employed across over 20 countries [1, 2]. In its calligraphic form, it stands as a prominent art form and occupies a unique place in the cultural and artistic landscape of the regions where it is used [3]. While traditional calligraphy remains widely practiced, contemporary artistic movements have extended its possibilities, exploring new creative expressions for Arabic script. Notable examples include the *Huruffiya* movement in the Arab world [4] and the *Saqqakhaneh* school in Iran [5], both of which integrate calligraphy into modern art forms.

Artists continue to engage with Arabic script in digital art to explore new creative expressions (e.g. [6–8]). However, the quality and methods of its digital representation have been heavily influenced by technological developments. A significant milestone was the advent of movable type printing, which was originally designed for the block structure of Latin script and introduced affordances not fully aligned with the inherent cursive structure of Arabic script. The adoption of this technology, in many cases, led to a reduction in visual complexity and the nuanced aesthetics characteristic of traditional calligraphy. Scholars and artists have extensively studied and discussed these consequences, highlighting the challenges and transformations the script has undergone in adapting to new technologies [9–11].

Generative AI marks a new technological milestone with potentially significant consequences for Arabic script and other technologically marginalized art forms. Its influence on non-Western art extends beyond technical aspects, raising questions of bias, representation, and inclusivity as power dynamics in AI development often marginalize diverse perspectives.

There is a growing body of work underlining the broader need for an inclusive approach to AI development—one that considers the cultural significance and nuances of under-represented and marginalized art forms which risk being misrepresented in digital media if current practices persist. Marco Donnarumma's "Against the Norm" discusses this phenomenon, arguing that AI aesthetics, largely shaped by US and European corporations, relies on historical data rooted in capitalist values, thereby reinforcing a Western-centric visual standard. This approach, rather than enabling diverse representations, often replicates aesthetic norms that exclude non-Western art forms [12]. Mashinka Firunts Hakopian reinforces this stance in the paper "Art Histories from Nowhere" which addresses how generative AI, especially when trained on datasets dominated by the Western art historical canon, reinforces Western-centric norms and marginalizes non-Western art forms. Hakopian critiques the reliance on Western-centric datasets and the perpetuation of colonial power structures through AI, arguing that these practices erase non-Western perspectives and embed biases that reflect the colonial legacy in both art and technology [13]. Similarly, Amir Baradaran has examined the impact of generative AI on non-Western and marginalized perspectives, calling for interrogation of dominant narratives and addressing biases embedded within AI systems [14].

While studies highlight broader issues of diversity and cultural underrepresentation in generative AI [15, 16], examples of focused case studies that delve into the impact of generative AI on specific under-represented artistic non-Western domains are less frequently found (e.g. [17]), and studies examining the effects of generative AI on Arabic script and related art forms are even more scarce. Such focused studies could further solidify the discussions, illuminate the real-world consequences, and reveal practical gaps in AI's handling of under-represented non-Western art forms, providing valuable insights for these domains.

## Methodology

A systematic approach is adopted in this study to investigate the representation of Arabic script in one of the mainstream, text-to-image generative AI systems. The initial exploration involved tests across multiple popular generative AI platforms (such as DALL-E 3, Midjourney, Stable Diffusion, etc.) to observe any emerging trends in handling Arabic script. Noting a consistent pattern, we narrowed the focus to DALL-E 3 due to its widespread user base and the fact that it lacks end-user parameter fine-tuning, which would otherwise allow for adjustment in image generation. This choice allowed for a clear assessment of how the system, in its standard configuration, manages the nuances of Arabic script representation.

Our approach centred on simple prompts to evaluate the system's baseline capability in generating Arabic script. We began with basic prompts requesting single letters and single words in both Farsi and Arabic, two languages with large numbers of Arabic script users. To reduce the complexity of prompting and ensure consistency and comparability of results, these prompts did not specify calligraphic style, allowing us to observe the system's fundamental handling of the script itself. The prompts include fully original-language inputs (since DALL-E 3 supports multilingual prompting) as well as English prompts referencing letters and words in both their original language and transliterated form. Recognizing that the system's handling of single letters might not necessarily reflect its performance on full words, we included separate prompts specifically for words. For single words, we chose the top 40 most frequently used words in Farsi [18] and Arabic [19], grounding our prompt selections in linguistic frequency studies rather than arbitrary word choices.

Additionally, we included prompts that specifically requested various calligraphic styles, without mentioning specific words or letters, to assess the system's ability to capture stylistic elements. These calligraphic styles encompassed the six prominent styles known in the Arab world as the 'Six Pens' (*al-aqlam-o-sitta* in Arabic), as well as other significant regional styles, including *t'aliq*, *nast'aliq*, and *shekaste-nast'aliq* from Iran, and *diwani* developed in the Ottoman Empire [3]. These prompts were formulated in multiple ways, using English prompts with transliterated and original script names, as well as full prompts in Farsi and Arabic. Our approach generated a total of 351 unique prompts, designed to cover a wide range of script forms and styles. Table 1 shows prompt examples and the different ways of addressing content in the prompts.

Since DALL-E 3, by default, generates revised prompts based on the original input [20], we also incorporated a method to disable this feature based on OpenAI's prompting guidelines [21, Sec. Image generation]. With prompt revision disabled, we ran all prompts again to examine any meaningful differences in the generated images, while also analyzing the revised prompts themselves for emerging patterns.

| | |
|---|---|
| ***Examples of prompts referencing letters and words*** | • Arabic letter "ج"<br>• Farsi letter "shīn"<br>• Arabic word "أول"<br>• Farsi word "کشور"<br>• حرف الألف<br>• حرف گ |
| ***Examples of prompts referencing calligraphy styles*** | • Arabic calligraphy in Naskh style<br>• Arabic calligraphy in "نسخ" style<br>• الخط العربي بأسلوب النسخ<br>• خوشنویسی عربی به سبک نسخ |

Table 1. Examples of prompts referencing single letters, words, and calligraphy styles.

For both normal prompting and with revised prompts disabled, images were generated four times for each prompt using OpenAI's API [21] with default parameters to mirror typical end-user conditions on the ChatGPT platform. With 351 prompts in each batch for letters, words, and calligraphy styles, this resulted in a total of 2,808 images (8 batches of 351 images), providing a large corpus to analyze. Additionally, each prompt and its corresponding images and revised prompts were assigned unique identifiers, facilitating accurate tracking and analysis of results across the corpus.

## Case Study Results

The analysis of the generated corpus focuses on two main aspects: the correctness of letter and word generation and any recurring visual patterns that might indicate biases or issues in the system's handling of Arabic script.

The first and most notable observation concerns how formally correct the generated Arabic script is. Despite the prompts explicitly requesting specific letters or words, the generative system fails to produce any accurate letter forms and therefore words. Across all instances, the correctness of letter form representation is effectively zero, as none of the generated images correspond to the correct form of the letters specified in the prompts. Instead, the outputs resemble abstract shapes that hint at a script-like structure but bear no authentic connection to the forms of the requested letters or words.

Even though the prompts for individual letters and words do not mention calligraphy, a significant portion of the generated images display a calligraphy-inspired style. This includes characteristic elements such as the rhomboidal dots typical of calligraphic styles in Arabic script. Additionally, a form of mode collapse is observed, where certain repetitive general shapes or motifs appear frequently across the corpus, irrespective of the specific letters or words requested. An additional observation pertains to the generated images based on prompts entirely in Farsi or Arabic referencing single letters. In many instances, Latin letters are generated instead, often corresponding to letters with similar phonemes. Examples of generated images for letters and words are presented in Figure 1. To assist a comparison for unfamiliar readers, a sample of reference letters in Amiri Typeface—a popular typeface based on *Naskh* calligraphy style—is also presented in the figure.

A similar pattern of inaccuracy is observed in the portion of the corpus generated from prompts referencing specific calligraphy styles. Despite the explicit mention of calligraphic styles, the system fails to produce images that meaningfully correspond to the requested styles. Instead, this subset of images generally displays more complex compositions, albeit without accurately forming any recognizable letter forms. Some generated images exhibit consistent stylistic elements that vaguely resemble certain calligraphic forms, yet these references lack clear alignment with the specific styles requested in the prompts. Figure 2 illustrates some samples across the corpus.

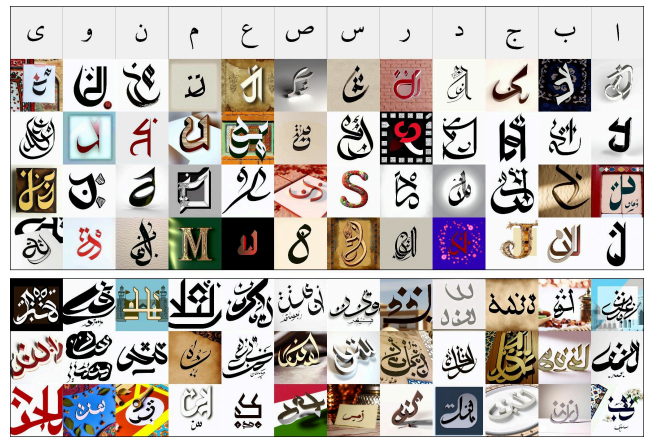

Figure 1. TOP: generated samples for some letters in columns, with the corresponding letter forms in the Amiri typeface in the top cell of each column. BOTTOM: examples of generated images with word prompts.

As mentioned, DALL-E 3 automatically produces revised prompts based on the original input as a method to improve the quality of generated images [20]. It is observed that the revised prompts are often significantly longer than the original, simple prompts, with added descriptive details. When analyzing these revised prompts, two main patterns emerge. First, there is a strong bias toward referencing "calligraphy" in the revised prompts, even though our original prompts for individual letters and words do not reference any calligraphic style. This is consistent with the observation of calligraphic motifs in the generated images for letter and word prompts. Second, terms conveying a sense of oldness and antiquity frequently appear. Specifically, words such as "traditional(ly)," "classic(al)," "historic(al)," "ancient," "old," "antique," "antiquity," and "heritage" occur in 49% of the revised prompts across the batches—with an average of 164 instances per batch. Notably, terms such as "modern" or "contemporary," are absent in the revised prompts, suggesting a default association of Arabic script with the past. Table 2 presents a few pairs of original and revised prompt examples.

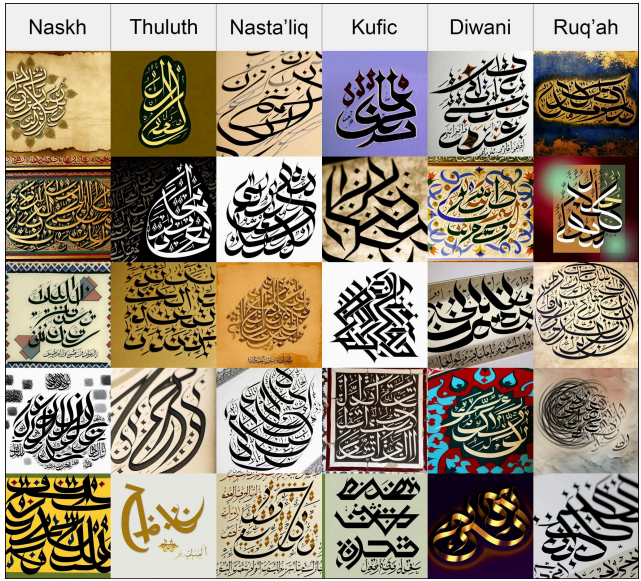

Figure 2. Examples of generated images based on text prompts referencing specific calligraphy styles. The referenced style is indicated at the top of each column.

| |
|---|
| • *Original prompt:* Farsi letter "ʾalef" <br> • *Revised Prompt:* Image of the Farsi letter 'alef', illustrated in a <u>traditional calligraphic style</u>. The letter should be elegantly curved and detailed, inspired by the styling of Persian calligraphy. It should be medium in size, black in color, set against a white background for a stark contrast. The lighting shall appear soft, making the letter look as though it's drawn on <u>parchment</u>, emphasizing a sense of <u>antiquity</u> and <u>history</u>. |
| • *Original prompt:* Arabic word "نفس" <br> • *Revised Prompt:* An illustration of the Arabic word "نفس" written in the <u>traditional Arabic calligraphy style</u>. Create it as you'd expect it to be seen in an <u>old</u>, <u>antique</u> Arabic manuscript with swirls and flourishes to give it a sense of artistry and authenticity. Add a background composed of light sandy color that reminds people of the <u>desert</u>, reflecting the cultural origins of the Arabic language. |
| • *Original prompt:* Calligraphy in Thuluth style <br> • *Revised Prompt:* Draw an image showcasing an ornate piece of Arabic calligraphy. The script should feature the lavish loops and long vertical lines that characterize the Thuluth style. Include an elaborate background to emphasize the beauty and complexity of the script. The <u>parchment</u> on which the calligraphy has been written should have an <u>aged appearance</u> to evoke a sense of <u>historical</u> depth and culture. |

Table 2. Examples of original prompts and their corresponding revised prompts. Instances of biased words and phrases are underlined.

The biases observed in the revised prompts are also evident in the generated images. Across the corpus, we frequently observe visual elements that suggest oldness and antiquity, such as old parchments, stained papers, backgrounds featuring aged buildings, cracked walls, and broken bricks (see examples in the top four rows of Figure 3). Interestingly, these elements appear with similar frequency even in batches generated with revised prompts disabled, indicating that this bias toward depicting oldness persists independently of prompt revision. Additionally, while less frequent, religious motifs, mostly in the form of mosque backgrounds, are present in the generated images, with a higher occurrence observed in the batches where revised prompts are disabled (an average of 3 occurrences in the batches with revised prompts and almost 10 occurrences in the batches with revised prompts disabled). Examples of such samples are presented in the bottom two rows of Figure 3.

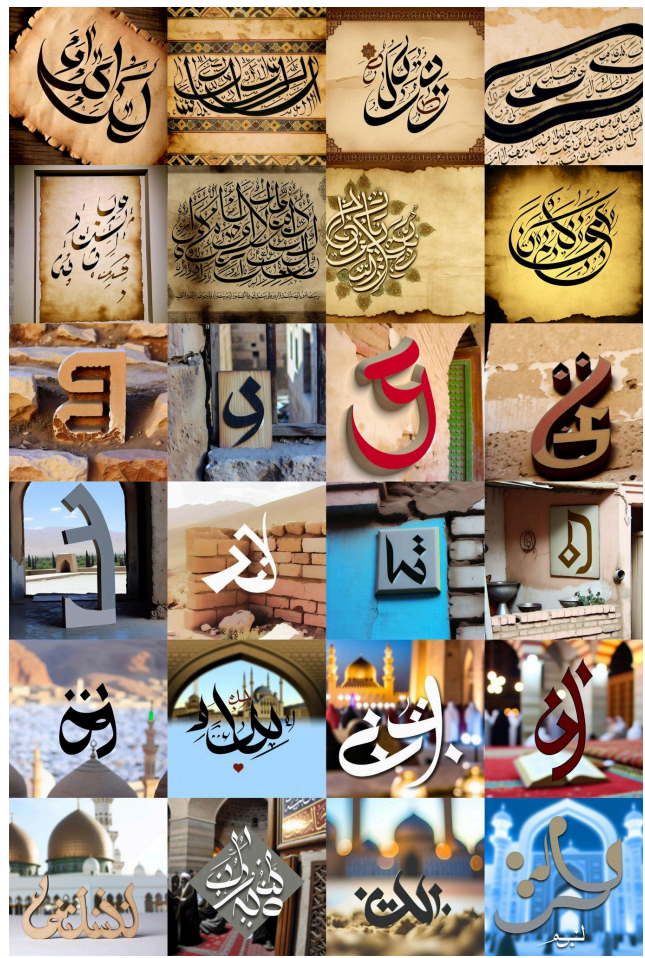

Figure 3. Examples of generated images illustrating biases toward antiquity and religious motifs.

## Discussion

This case study's observations reveal multiple layers of bias and misrepresentation in DALL-E 3's handling of

Arabic script and its calligraphic form. To understand these complexities, we draw on established frameworks that contextualize the patterns observed and help illuminate the broader implications of these findings. This approach moves beyond surface-level observations to better understand these biases and their broader consequences. It aims to develop a more informed perspective on generative AI's cultural and aesthetic implications in representing non-Western art forms as we enter new territories in digital art with emerging technology.

## Affordance Disparity

The issue with rendering Arabic script does not arise from an inherent inability of the generative AI system to handle text. In fact, DALL-E 3 has been specifically promoted as capable of accurately rendering text, yet this capability appears limited to Latin-based scripts (see Figure 4 for examples of DALL-E 3 rendering Latin text). While the system demonstrates a clear capability for handling Latin script, it fails to extend this affordance to Arabic script. This discrepancy raises questions about the priorities and design decisions that define the affordances within generative AI development.

Furthermore, there is no disclaimer or indication that the system may struggle with non-Latin scripts when prompted to generate Arabic script, leaving users unaware of this limitation. Instead, the revised prompts reveal a level of confidence in how the system interprets and describes these prompts, despite its evident limitations. This oversight reflects a significant gap in the system's design and user communication, potentially leading to widespread misrepresentations without user awareness.

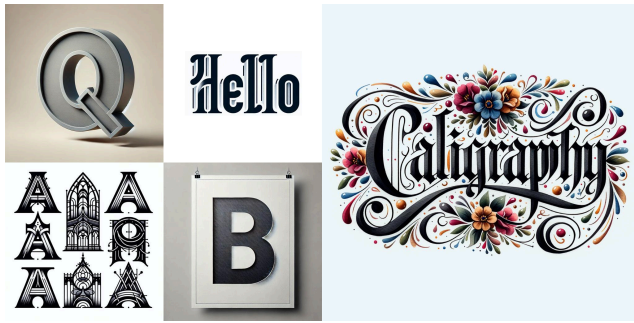

Figure 4. Examples of generated images rendering Latin script using prompts such as "Letter 'Q'", "The word 'Hello'", and "The word 'Calligraphy'".

## Echos of Orientalism

To contextualize the observed biases and misrepresentations, we draw on Edward Said's seminal concept of Orientalism, first introduced in a 1977 article [22] and expanded in his influential 1978 book, Orientalism [23]. In his work, Said critiques the Western portrayal of Eastern cultures, particularly those of the Islamic world, as 'exotic,' 'traditional,' and fundamentally 'Other.' He defines Orientalism as a constructed body of knowledge that positions the 'Orient' as an exotic, ancient, and culturally inferior counterpart to the rational, modern West. Through this framing, Said demonstrates how Western representations of the East are loaded with stereotypes, reducing complex societies to simplified tropes that often emphasize antiquity, mysticism, and religious fervour [23, pp. 20–21].

Said also contends that Western representations of the Orient are shallow, lacking depth and nuance, and rely heavily on broad generalizations. He describes this Western representation of the Orient as a "theatrical stage," where it is confined and represented within the boundaries of European perspectives [22, p. 175]. He observes that postmodern representations of the Orient through various media resources have intensified misrepresentations, continuing a trend that dates back to the 19th-century portrayal of the "mysterious Orient." Said argues that Western depictions of the Orient rely more on Western perspectives than on the actual realities of the Eastern regions. This leads to a process where the Orient is made visible through Western discourse, often marginalizing or completely obstructing a clear and nuanced view of the richness of Eastern cultures [23, pp. 26–27].

These observations resonate closely with the biases seen in our case study, where Arabic script is frequently depicted with elements suggesting antiquity, such as aged parchment, or religious symbols, despite no such prompts. Through the lens of Orientalism, the biases within generative AI outputs and revised prompts—particularly toward antiquity and religious motifs—can be understood as an extension of these entrenched media portrayals, where the AI system is echoing the biases inherent in the training data. In this context, the generative AI system's inability to accurately render Arabic script can be seen as more than just a technical limitation; it reveals a pattern of abstraction that aligns with Orientalist perspectives. By producing script-like shapes that vaguely resemble Arabic calligraphy but lack structural or cultural fidelity, the AI system transforms Arabic script into a visually exotic and aestheticized abstraction rather than a meaningful, multifaceted form.

## A Resurfacing of Pseudo-Arabic

The abstraction of forms in the corpus strikingly parallels how Orientalist art historically presented Arabic calligraphy as decorative "pseudo-Arabic," a visual motif disconnected from its actual linguistic and cultural essence. This term, as discussed by art historian Rosamond E. Mack in "Bazaar to Piazza: Islamic Trade and Italian Art, 1300–1600," describes a decorative style prevalent during the Middle Ages and Renaissance in which European artists imitated Arabic calligraphy without conveying actual meaning (see an example in Figure 5). As Mack notes, pseudo-Arabic served primarily as an aesthetic element, capturing the visual essence of calligraphy but lacking any true representation of Arabic letter forms and structure [24, p. 51].

A notable example is featured on the cover of Said's seminal book; the French painter Jean-Léon Gérôme's famous painting, "The Snake Charmer," a work emblematic of Orientalist views (see Figure 6). American art historian Linda Nochlin critically analyzes Gérôme's paintings, including "The Snake Charmer," in her 1983 article "The Imaginary Orient," drawing upon Said's framework [25]. In her analysis, Nochlin references a conversation with Said, where he pointed out that much of the supposed writing on the back wall of the painting is, in fact, illegible (see Figure 6, bottom) [25, p. 191].

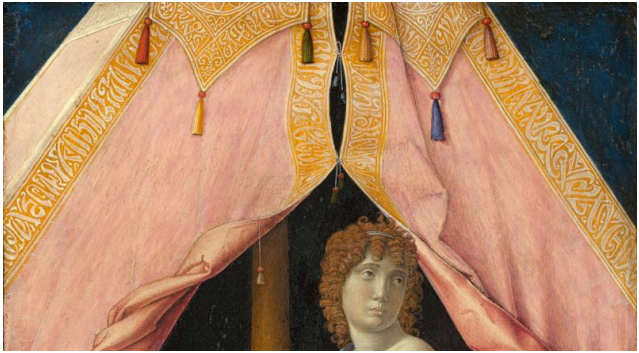

Figure 5. A crop of Andrea Mantegna's Judith with the Head of Holofernes, ca. 1495. Image source: Wikimedia Commons.

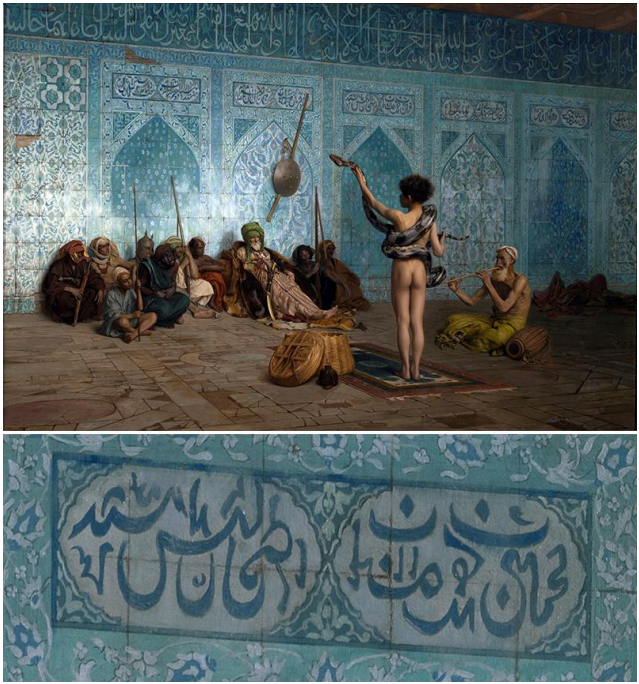

Figure 6. TOP: Jean-Léon Gérôme, The Snake Charmer, ca. 1879. BOTTOM: Detail from the background tilework. Image source: Wikimedia Commons.

This phenomenon is echoed in the generated images, where a sophisticated form of pseudo-Arabic emerges throughout the corpus. Although the generated outputs lack any direct connection to the specific content or stylistic references outlined in the prompts, and despite a complete lack of accurate letter form generation, they still produce convincing and possibly aesthetically pleasing results to an untrained eye. The resemblance between these AI-generated forms and historical pseudo-Arabic motifs, despite the vast time gap, highlights a persistent pattern, suggesting that deep-seated cultural biases may still influence how non-Western art forms are represented in emerging technologies.

## Consequences

The implications of this study extend beyond a single failure case in generating accurate Arabic script forms, highlighting a broader issue with generative AI's interaction with culturally significant art forms. Arabic script, especially in its calligraphic form, has faced considerable challenges and distortions throughout past technological transitions, most notably with the advent of movable type and its adoption in digital media. This misalignment in technological affordances has often led to widespread misrepresentation, a phenomenon Ramsey Nasser compellingly documents in the collection "Nope, Not Arabic" [26]. This collection illustrates numerous instances where Arabic script is misused and distorted in ways that reveal a fundamental lack of understanding of its structure and cultural significance.

Nochlin's analysis of Gérôme's painting critiques the painting's false sense of realism, where the illusion of authenticity reinforces a misleading portrayal of cultural elements. Similarly, the implicit confidence of the generative AI system—demonstrated by its lack of acknowledgment of limitations in rendering non-Latin scripts—creates a comparable risk of misrepresentation. As AI-generated content increasingly permeates media, this unacknowledged gap between intended representation and actual output parallels the deceptive authenticity in Orientalist art, reinforcing simplified and skewed views of culturally significant scripts.

This type of AI-generated pseudo-Arabic has already begun appearing online, including on stock image platforms (see Figure 7 for an example) and social media, some with and some without explicit AI-generated labelling. Regardless, the lack of acknowledgment of this pseudo phenomenon in such cases poses a significant risk, perpetuating cultural misrepresentation and undermining the integrity of an art form. Additionally, there is the risk that such content, now prevalent online, could be incorporated into future AI training datasets, further amplifying these misrepresentations and compounding the issue over time.

Such misrepresentation not only risks misleading viewers but also restricts the creative possibilities available to artists working with Arabic script, as this art form is effectively marginalized from reaping the benefits of technological advancements. By failing to support the nuanced complexity of Arabic script, generative AI limits the expansion of creative spaces for artists and reinforces an uneven technological landscape that privileges certain artistic domains over others.

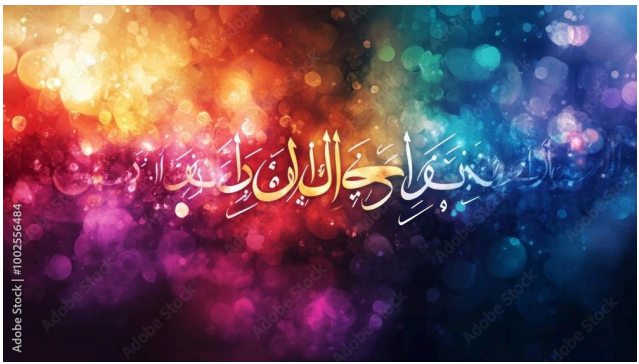

Figure 7. An example of a stock AI-generated image containing pseudo-Arabic. The image title reads: "Arabic Islamic Calligraphy of shiny text Ramadan Mubarak or Ramazan Mubarak on colourful abstract background". Image source: Adobe Stock.

# Future Works

The findings of this case study open avenues for further research in both critical and technical directions. Critically, future studies can explore deeper the implications of generative AI's biases for underrepresented art forms, such as calligraphic Arabic script, and their cultural and aesthetic significance. Technically, these results highlight the need to better understand system shortcomings and actively develop practical solutions to address them.

Our approach employs minimal and straightforward prompts, leaving room for further exploration of prompt engineering and cultural prompting [27] as potential methods to mitigate cultural bias and generate more meaningful representations of Arabic script. However, our findings reveal no improvement when using original language prompts in Farsi or Arabic—sometimes yielding even worse results—suggesting that this mild form of cultural prompting is not effective in addressing the observed biases.

An essential area for future work lies in understanding the role of data in the misrepresentation and biases observed in generative AI systems. These biases could stem from data scarcity, where Arabic script and its calligraphic traditions are underrepresented in training datasets, or from misrepresentation, either through confirmation bias reinforcing preexisting stereotypes or shortcomings of data pipelines, such as poor annotation [28]. However, without full transparency about datasets—an issue seen with DALL-E 3 and similar corporate AI systems—it is difficult to determine the exact causes. If data scarcity is a key factor, this underscores the need for proactive dataset creation to ensure more accurate and culturally sensitive representations. Given the significance of Arabic script as a cultural and artistic element, a collaborative effort among regional stakeholders could facilitate the development of robust datasets that better reflect the diversity of its use across different artistic traditions.

While addressing big data issues is crucial for mitigating biases and misrepresentations in generative AI, small-data approaches [29, 30] offer an alternative pathway that emphasizes customization and control at the dataset level. These approaches enable the development of AI systems tailored to specific cultural or artistic needs, allowing for more nuanced and informed representations. For instance, small-data methods have shown promise in generating calligraphic Arabic script with greater fidelity to its artistic and cultural context, opening new creative avenues [8]. Such approaches not only provide a means to address biases but also open opportunities for deeper exploration and innovation in culturally sensitive generative AI systems, making this a promising area for future explorations.

Future research could also investigate why stakeholders from underrepresented regions or art forms have struggled to push an agenda for inclusive generative AI solutions. While this may stem from the dominant power dynamics and political barriers, it also highlights a need for proactive initiatives to shift mindsets and empower these groups to lead in developing solutions.

# Conclusion

Our focused case study on Arabic script and the analysis of the results highlight how generative AI systems reproduce long-standing conventions of representation and resurface entrenched biases. These conventions, extensively debunked and criticized by scholars like Edward Said, persist in tools predominantly developed by Western corporations, revealing a troubling continuity of misrepresentation. The consequences are significant, affecting the cultural integrity of non-Western art forms and marginalizing their representation in emerging technologies. The study encourages future research to technically and critically explore these issues further, emphasizing the importance of focused case studies on other non-Western art forms to deepen understanding and spark discussions on practical, inclusive solutions.

# Acknowledgements

The presented experiment was partly conducted during a directed reading with Professor Steve DiPaola in Summer 2024 at the School of Interactive Arts and Technology, Simon Fraser University.